\documentclass[12pt,a4paper]{article}

\usepackage{amssymb}
\usepackage{epsf}

\begin{document}

\title{Velocity of electroweak bubble walls}
\author{ {\large Ariel M\'egevand\thanks{%
Member of CONICET, Argentina. E-mail address: megevand@mdp.edu.ar}~ and
Alejandro D. S\'anchez\thanks{%
Member of CONICET, Argentina. E-mail address: sanchez@mdp.edu.ar}} \\
{\normalsize \emph{IFIMAR (CONICET-UNMdP)}, }\\
{\normalsize \emph{Departamento de F\'{\i}sica, Facultad de Ciencias Exactas
y Naturales, UNMdP,} }\\
{\normalsize \emph{De\'an Funes 3350, (7600) Mar del Plata, Argentina} }}
\maketitle

\begin{abstract}
We study the velocity of bubble walls in the electroweak phase
transition. For several extensions of the Standard Model, we
estimate the friction and calculate the wall velocity, taking
into account the hydrodynamics. We find that deflagrations are
generally more likely than detonations. Nevertheless, for
models with extra bosons, which give a strongly first-order
phase transition, the deflagration velocity is in general quite
high, $0.1\lesssim v_w\lesssim 0.6$. Therefore, such phase
transitions may produce an important signal of gravitational
waves.  On the other hand, models with extra fermions which are
strongly coupled to the Higgs boson may provide a strongly
first-order phase transition and small velocities,
$10^{-2}\lesssim v_w\lesssim 10^{-1}$, as required by
electroweak baryogenesis.
\end{abstract}

\section{Introduction}

The electroweak phase transition may give rise to a variety of
cosmological relics such as the baryon asymmetry of the
universe, cosmic magnetic fields, inhomogeneities or
gravitational waves. To be observable, some of these relics
depend on the strength of the phase transition. This is the
case, e.g., of the baryon asymmetry of the Universe (BAU) and
gravitational waves (GWs). Both can be generated in the
electroweak phase transition, and both require a strongly
first-order phase transition. In a first-order phase
transition, bubbles of the stable broken-symmetry phase
nucleate at a temperature $ T_{n}$ below the critical
temperature $T_{c}$, and grow inside the supercooled symmetric
phase. The expansion of bubbles provides the required departure
from thermal equilibrium.

Gravitational waves  are generated by the collisions of bubble
walls and by the turbulence they produce \cite{kkt94,kmk02}. As
a consequence, bubble walls with higher velocities will
generate GWs of larger amplitudes. For electroweak baryogenesis
\cite{ckn93}, the violation of baryon number must be suppressed
in the broken-symmetry phase in order to avoid the wash-out of
the generated BAU. This puts a condition on the vacuum
expectation value (VEV) of the Higgs field in that phase,
namely, $\langle \phi \rangle /T\gtrsim 1$. This quantity is
the order parameter of the phase transition, and the above
condition implies that the phase transition must be strongly
first-order. The amplitude of the baryon asymmetry also depends
on the wall velocity. However, unlike the case of gravitational
waves, the BAU has a maximum for wall velocities in the range
$v_{w}\sim 10^{-2}-10^{-1}$, depending on the model
\cite{lmt92,ckn92}. For higher velocities the sphaleron
processes do not have enough time to produce baryons, whereas
for lower velocities equilibrium is restored and the generated
baryon number is washed-out. The low velocities required for
baryogenesis seem to be too small to generate observable
gravitational radiation.

For baryogenesis calculations the wall velocity is often
assumed to be of the form
\cite{lmt92,ckn92,mp95,js01,dlhll92,k92,m00}
\begin{equation}
v_{w}\approx \Delta p(T)/\eta ,  \label{vnr}
\end{equation}%
where $\Delta p$ is the net pressure acting on the wall, i.e.,
the pressure difference between the two phases, and $\eta $ is
a friction coefficient. This approximation corresponds to the
case of subsonic walls, which propagate as
\emph{deflagrations}. In the context of gravitational waves, on
the contrary, the wall is generally assumed to propagate as a
\emph{detonation}
\cite{kmk02,dgn02,cd06,kkgm08,hk08,amnr02,n04,gs07}.
Furthermore, a \emph{Jouguet} detonation is assumed, leading to
a simple expression for the velocity \cite{s82},
\begin{equation}
v_{w}=\frac{\sqrt{1/3}+\sqrt{\alpha ^{2}+2\alpha /3}}{1+\alpha },
\label{detojug}
\end{equation}%
which depends only on the parameter $\alpha =L/\rho
_{\mathrm{rad}}$, where $ L$ is the latent heat and $\rho
_{\mathrm{rad}}$ is the energy density of radiation. Equation
(\ref{detojug}) does not depend on the friction, and the
Jouguet velocity gives in fact a lower bound for the detonation
velocity \cite{l94,hkllm93}. In a recent paper \cite{ms09} we
investigated the wall velocity as a function of $L,$ $\eta ,$
and the supercooling temperature $T_{n}$. We found that Eq.
(\ref{vnr}) is a good approximation for nonrelativistic
velocities, while Eq. (\ref{detojug}) grossly underestimates
the actual detonation velocity.

The calculation of the friction coefficient $\eta $ is
involved. It depends on all the particle species that are
present in the plasma and their interactions. An accurate
computation of $\eta $ thus requires considering the details of
the particle content in each specific model (see e.g.
\cite{mp95,js01}). To compare different models it is more
workable to consider approximations which depend on a few
parameters. Such a simple approximation was obtained in Ref.
\cite{m04} using previous results
\cite{lmt92,mp95,js01,dlhll92,k92},
\begin{equation}
\eta _{\mathrm{th}}=\sum \frac{g_{i}h_{i}^{4}}{\Gamma _{i}/T}\left(
\frac{\log \chi _{i}}{2\pi ^{2}}\right) ^{2}\frac{\phi ^{2}\sigma }{T}
\label{etatth}
\end{equation}%
for particles with a thermal distribution, whereas infrared
bosons contribute a term \cite{m00}
\begin{equation}
\eta _{\mathrm{ir}}=\sum_{\mathrm{bosons}}\frac{g_{i}m_{D}^{2}T}{32\pi L_{w}}
\log \left( m_{i}\left( \phi\right) L_{w}\right) .  \label{etatir}
\end{equation}%
In these equations, $g_{i}$ is the number of degrees of freedom
(d.o.f.) of species $i$ with Higgs-dependent mass
$m_i=h_{i}\phi$, $\Gamma _{i}$ are interaction rates which are
typically $\lesssim 10^{-1}T$, $\chi _{i}=2$ for fermions and
$\chi _{i}=m_{i}\left( \phi \right) /T$ for bosons, $\sigma $
is the surface tension of the bubble wall, $m_{D}^{2}\sim
h_{i}^{2}T^{2}$ is the Debye mass squared, and $L_{w}$ is the
width of the bubble wall, $L_{w}\approx \phi ^{2}/\sigma $. The
derivation of Eqs. (\ref{etatth}) and (\ref{etatir}) involves
expanding the distribution functions to lowest order in
$m_{i}/T$, and thus they break down for $m_{i}/T\gg 1$. In
particular, for a squared mass of the form
$m_i^2=\mu_i^2+h_i^2\phi^2$, with large $\mu _{i}$, the
particle density in the symmetric phase will be suppressed by a
Boltzmann factor $\exp (-\mu_i /T)$, and this species will not
contribute to the friction.

In this paper we study the velocity of bubble walls in the
electroweak phase transition for several extensions of the
Standard Model (SM). For that aim, we modify Eqs.
(\ref{etatth}) and (\ref{etatir}) to take into account more
general masses $m_i(\phi)$. We also take into account the
effects of hydrodynamics in order to consider both deflagration
and detonation solutions. We consider several extensions of the
SM, including the Minimal Supersymmetric Standard Model and
extensions with singlet scalars and with heavy fermions. We
also investigate the effect of cubic terms in the tree level
potential.

The plan is the following. In section \ref{dynamics} we review
the dynamics of a first-order electroweak phase transition. In
section \ref{micro} we find an approximation for the friction
coefficient which is valid for large as well as for small
values of $ m_i(\phi)/T$. In section \ref{veloc} we write down
the equations for the wall velocity, which we solve
numerically. The result depends on several parameters, namely,
the critical temperature $T_c$, the nucleation temperature
$T_n$, the latent heat $L$, and the friction coefficient
$\eta$. We compute these parameters for several models in
section \ref{modelos}, and we calculate the wall velocity.
Finally, in section \ref{conseq} we discuss the implications of
our results for baryogenesis and gravitational wave production
in the electroweak phase transition. Our conclusions are
summarized in section \ref{conclu}.

\section{Dynamics of the electroweak phase transition \label{dynamics}}

In the SM, the electroweak phase transition is only a smooth
crossover. However, many extensions of the model give a
first-order phase transition. For simplicity we shall consider
models with a single Higgs field, or models for which
considering a single Higgs provides a good approximation. Thus,
our theory will consist of a tree-level potential
\begin{equation}
V_{0}\left( \phi \right) =-m^{2}\phi ^{2}+\frac{\lambda }{4}\phi ^{4},
\label{v0}
\end{equation}
for a scalar field $\phi $ (the background Higgs field, defined
by $\langle H^{0}\rangle \equiv \phi /\sqrt{2}$). The vacuum
expectation value of the Higgs is given by $v=\sqrt{ 2/\lambda
}m=246GeV$, and $\lambda $ fixes the Higgs mass, $
m_{H}^{2}=2\lambda v^{2}$. Imposing the renormalization
conditions that the minimum of the potential and the mass of
$\phi $ do not change with respect to their tree-level values
\cite{ah92}, the one-loop zero-temperature potential is given
by $V\left( \phi \right) =V_{0}\left( \phi \right) +V_{1}\left(
\phi \right) $, with
\begin{equation}
V_{1}\left( \phi \right) =\sum_{i}\frac{\pm g_{i}}{64\pi ^{2}}\,\left[
m_{i}^{4}(\phi )\left( \log \left( \frac{m_{i}^{2}(\phi )}{m_{i}^{2}(v)}
\right) -\frac{3}{2}\right) +2m_{i}^{2}(\phi )m_{i}^{2}(v)\right] +c,
\label{v1loop}
\end{equation}%
where $g_{i}$ is the number of d.o.f. of each particle species,
$m_{i}\left( \phi \right) $ is the $\phi $-dependent mass, and
the upper and lower signs correspond to bosons and fermions,
respectively. We have added a constant $c$ such that $V\left(
v\right) =0$, so that the energy density vanishes in the true
vacuum at zero temperature. In the symmetric phase we will have
a false vacuum energy density, given by $\rho _{\Lambda
}=V\left( 0\right) $, which contributes to the Hubble rate
during the phase transition. For particle masses of the form
$m_{i}^2=h_{i}^{2}\phi ^{2}+\mu _{i}^{2}$, we have
\begin{equation}
\rho _{\Lambda }=\left[ \lambda +\sum_{i}\frac{\mp g_{i}}{32\pi ^{2}}\left(
h_{i}^{4}-2h_{i}^{2}\left( \frac{\mu _{i}}{v}\right) ^{2}-2\left( \frac{\mu
_{i}}{v}\right) ^{4}\log \left( \frac{\mu _{i}}{v}\right) ^{2}\right) \right]
\frac{v^{4}}{4}.
\end{equation}%
The free energy is given by the finite-temperature effective
potential. To one-loop order, including the resummed daisy
diagrams, we have
\begin{equation}
\mathcal{F}=V_{0}\left( \phi \right) +V_{1}\left( \phi \right) +
\mathcal{F}_{1}(\phi ,T),  \label{ftot}
\end{equation}%
where the finite-temperature corrections are given by
\cite{quiros}
\begin{eqnarray}
\mathcal{F}_{1}(\phi ,T) &=&\sum_{i}\pm \frac{g_{i}T^{4}}{2\pi ^{2}}
\int_{0}^{\infty }dx\,x^{2}\log \left[ 1\mp \exp \left( -\sqrt{
x^{2}+m_{i}^{2}\left( \phi \right) /T^{2}}\right) \right]  \label{f1loop} \nonumber \\
&&+\sum_{bosons}\frac{g_{i}T}{12\pi }\left[ m_{i}^{3}\left( \phi \right) -
\mathcal{M}_{i}^{3}\left( \phi \right) \right] ,
\end{eqnarray}%
where the upper sign stands for bosons, the lower sign stands for fermions,
and $\mathcal{M}_{i}^{2}\left( \phi \right) =m_{i}^{2}\left( \phi \right)
+\Pi _{i}\left( T\right) $, where $\Pi _{i}\left( T\right) $ are the thermal
masses. The last term receives contributions from all the bosonic species
except the transverse polarizations of the gauge bosons.

At high temperature the symmetry is restored, and in a certain
range of temperatures, the symmetric minimum $\phi =0$ coexists
with a symmetry-breaking minimum $\phi _{m}(T)$. The free
energy density of the unbroken-symmetry phase is given by
$\mathcal{F} _{u}(T)=\mathcal{F}(0,T)$, whereas that of the
broken-symmetry phase is given by
$\mathcal{F}_{b}(T)=\mathcal{F}(\phi _{m}(T),T)$. The critical
temperature is that for which $\mathcal{F}
_{u}(T_{c})=\mathcal{F}_{b}(T_{c})$. The energy density in each
phase is given by $\rho \left( T\right)
=\mathcal{F}(T)-T\mathcal{F}^{\prime }(T)$ (a prime indicates
derivative of a function with respect to its variable), and the
latent heat $L\equiv \rho _{u}\left( T_{c}\right) -\rho
_{b}\left( T_{c}\right) $ is given by
\begin{equation}
L=T_{c}\left( \mathcal{F}_{b}^{\prime }(T_{c})-\mathcal{F}_{u}^{\prime
}(T_{c})\right) .  \label{ele}
\end{equation}%
We define the thermal energy density $\tilde{\rho}_{u}$ by
subtracting the vacuum energy density, $\tilde{\rho}_{u}=\rho
_{u}-\rho _{\Lambda }$. In general, we have
$\tilde{\rho}_{u}\approx \pi ^{2}g_{\ast }T^{4}/30$, where
$g_{\ast }$ is the number of relativistic d.o.f. The
hydrodynamics of the bubble wall will depend on the parameters
$\alpha _{c}=L/\tilde{\rho} _{u}\left( T_{c}\right) $ and
$\alpha _{n}=L/\tilde{\rho}_{u}\left( T_{n}\right) ,$ where
$T_{n}$ is the nucleation temperature.

The nucleation of bubbles \cite{c77,nucl} is governed by the
three-dimensional instanton action
\begin{equation}
S_{3}=4\pi \int_{0}^{\infty }r^{2}dr\left[ \frac{1}{2}\left( \frac{d\phi }{dr
}\right) ^{2}+V_T\left( \phi \left( r\right) \right) \right] ,  \label{s3}
\end{equation}
where
\begin{equation}
V_T(\phi )\equiv \mathcal{F}(\phi ,T)-\mathcal{F}(0,T).  \label{effpot}
\end{equation}
The bounce solution of this action, which is obtained by
extremizing $S_{3},$ gives the radial configuration of the
nucleated bubble, assumed to be spherically symmetric. The
action of the bounce coincides with the free energy of a
critical bubble (i.e., a bubble in unstable equilibrium between
expansion and contraction). This solution obeys the equation
\begin{equation}
\frac{d^{2}\phi }{dr^{2}}+\frac{2}{r}\frac{d\phi }{dr}=V_T^{\prime }\left(
\phi \right)  \label{eqprofile}
\end{equation}%
with boundary conditions
\begin{equation}
\frac{d\phi }{dr}\left( 0\right) =0,\ \lim_{r\rightarrow \infty }\phi \left(
r\right) =0.
\end{equation}%
We will solve Eq. (\ref{eqprofile}) iteratively by the
overshoot-undershoot method\footnote{See Ref. \cite{ms08} for
details.}. The thermal tunneling probability for bubble nucleation
per unit volume and time is \cite{nucl}
\begin{equation}
\Gamma_n \left( T\right) \simeq A\left( T\right) e^{-S_{3}\left( T\right) /T},
\label{gamma}
\end{equation}%
with $A\left( T\right) =\left[ S_{3}\left( T\right) /(2\pi
T)\right] ^{3/2}$. The nucleation time $t_{n}$ is defined as
that at which the probability of finding a bubble in a causal
volume is 1,
\begin{equation}
\int_{t_{c}}^{t_{n}}dt\Gamma_n \left( T\right) V_{c}=1,  \label{intnucl}
\end{equation}%
where $t_{c}$ is the time at which the Universe reached the critical
temperature $T_{c}$ and, in the radiation-dominated era, the causal volume
is given by $V_{c}\sim \left( 2t\right) ^{3}$. The time-temperature relation
is given by
\begin{equation}
dT/dt=-HT,  \label{tT}
\end{equation}%
where $H$ is the expansion rate, $H=\sqrt{8\pi G\rho
_{u}(T)/3}$. Here, $G$ is Newton's constant. Using Eq.
(\ref{tT}) we can solve  Eq. (\ref{intnucl}) for the
temperature $T_{n}$ at which the first bubbles are nucleated.
If $\rho _{u}\approx \tilde{\rho}_{u}\approx \pi ^{2}g_{\ast
}T^{4}/30$, then the time-temperature relation is given by the
usual expression $t=\xi M_{P}/T^{2} $, where $M_{P}$ is the
Planck mass and $\xi =\sqrt{45/(16\pi ^{3}g_{\ast })} $.

It is useful to consider the profile $\phi \left( r\right) $ of
the critical bubble at $T\approx T_{c}$. At the critical
temperature, the radius of the nucleated bubble diverges (and
the nucleation rate vanishes). Hence, for $T\approx T_{c}$ the
second term in Eq. (\ref{eqprofile}) can be neglected, since
the wall is much thinner than the radius. Thus, one obtains
\begin{equation}
d\phi/dr  =-\sqrt{2V_T\left( \phi \left( r\right) \right)
}.  \label{thin}
\end{equation}%
Within this approximation, the wall is planar and its surface
tension $ \sigma \equiv \int_{-\infty }^{+\infty }\left( d\phi
/dr \right) ^{2}dr$ is given by
\begin{equation}
\sigma =\int_{0}^{\phi _{c}}\sqrt{2V_T\left( \phi \right) }d\phi ,
\label{sigma}
\end{equation}
where $\phi _{c}\equiv\phi _{m}\left( T_{c}\right) $. We can
also invert relation (\ref{thin}) to obtain the wall width
$L_w$. If we define, e.g., $L_{w}=r\left( \phi =0.1\phi
_{c}\right) -r\left( \phi =0.9\phi _{c}\right) $, we have
\begin{equation}
L_{w}=\int_{0.1\phi _{c}}^{0.9\phi _{c}}d\phi /\sqrt{2V_T\left( \phi \right) }.
\label{lw}
\end{equation}%
Roughly, $\sigma$ and $L_w$ are related by $\sigma \sim \phi
_{c}^{2}/L_{w}$, and the wall width is given by $L_{w}\sim
V_T^{\prime \prime }\left( 0\right) ^{-1/2}$.

\section{Microphysics \label{micro}}

According to kinetic theory, for a planar wall in stationary motion along
the $z$ direction, the friction force per unit area is given by \cite{mp95}
\begin{equation}
\mathrm{friction}=\sum g_{i}\int_{-\infty }^{+\infty }dz\int
\frac{d^{3}p}{\left( 2\pi \right) ^{3}}\frac{dE}{dm^{2}} \frac{dm_{i}^{2}}{d\phi }
\frac{d\phi }{dz}\delta f_{i},  \label{fric}
\end{equation}%
where $E=\sqrt{p^{2}+m^{2}}$ and $\delta f_{i}$ is the departure from the
equilibrium distribution $f_{0}(E_{i}/T)$ for each particle species, with
\begin{equation}
f_{0}\left( x\right) =\frac{1}{e^{x}\pm 1}.
\end{equation}%
The kinetic description is valid for particles with $p\gg
L_{w}^{-1}$, for which the background field varies slowly and
the semiclassical (WKB) approximation is valid. Since in
general $L_{w}^{-1}\ll T$, this condition is satisfied for all
but the most infrared particles \cite{mp95}. As usual, we will
assume that the friction is proportional to the wall velocity,
\begin{equation}
\mathrm{friction}=\eta v_{w},  \label{fricnr}
\end{equation}%
where the friction coefficient $\eta $ is obtained by
considering Eq. (\ref{fric}) to linear order in $v_{w}$. The
deviations $\delta f_{i}$ can in principle be calculated by
considering the Boltzmann equation for the distribution
functions \cite{lmt92,mp95,js01,dlhll92,k92}. However, infrared
excitations of bosonic fields  should be treated classically
and undergo overdamped evolution \cite{m00,mt97,asy97}. We
shall refer to these fields as ``infrared bosons", whereas we
shall call ``thermal particles" those which obey the Boltzmann
equation.

\subsection{Thermal particles}

We begin by considering the case of thermal particles. It is
usual to employ the ansatz $f=f_{0}\left( E/T-\mu /T+E\delta
T/T^{2}+p_{z}v/T\right) $ for the distribution functions
\cite{mp95,js01}. In that case, a system of equations for $\mu
$, $\delta T$ and $v$ for each particle species can be derived
from the Boltzmann equation
\begin{equation}
\left[ \partial _{t}+\left( \partial _{p_{z}}E\right) \partial _{z}-\left(
\partial _{z}E\right) \partial _{p_{z}}\right] f=-C\left[ f\right] ,
\label{boltzmann}
\end{equation}%
where $C\left[ f\right] $ is the collision term. We need to
simplify further the problem in order to obtain a simple
analytical expression which can be applied to different models.
Therefore, we shall use the ansatz $ f=f_{0}\left( E/T-\delta
\right) $, which is equivalent to considering only the term
$\mu /T$. Hence, the deviation from $f_{0}\left( E/T\right) $
is $ \delta f=-f_{0}^{\prime }\left( E/T\right) \delta $, and
the equation for $ \delta $ is obtained by linearizing the
Boltzmann equation. Assuming stationary motion and making a
momentum integration, one obtains, for each particle species
(see Ref. \cite{m04} for details),
\begin{equation}
c_{2}v_{w}\frac{d\delta}{dz} -\Gamma \delta =\frac{c_{1}v_{w}}{2T^{2}}
\frac{dm^{2}}{dz},
\label{fluideq}
\end{equation}%
where $\Gamma $ is an interaction rate arising from the collision integrals,
and the coefficients $c_{1}$ and $c_{2}$ are defined by
\begin{equation}
c_{1}\equiv -\frac{1}{T^{2}}\int \frac{d^{3}p}{\left( 2\pi \right) ^{3}E}
f_{0}^{\prime }\left( E/T\right) ,\quad c_{2}\equiv -\frac{1}{T^{3}}\int
\frac{d^{3}p}{\left( 2\pi \right) ^{3}}f_{0}^{\prime }\left( E/T\right) .
\nonumber
\end{equation}%
It is out of the scope of this work to calculate the collision
integrals, which depend on the particle content of each model.
Numerically, the rates $\Gamma $ are $\sim 10^{-2}T$
\cite{mp95}, and we shall set $\Gamma/T =5\times 10^{-2}$. In
the thick wall limit, the first term in Eq. (\ref{fluideq}) can
be neglected, and we obtain
\begin{equation}
\delta =-\frac{c_{1}v_{w}}{2T^{2}\Gamma}\frac{dm^{2}}{dz}.  \label{delta}
\end{equation}%
We now insert $\delta f=-f_{0}^{\prime }\left( E/T\right)
\delta  $ in Eq.~(\ref{fric}), with $\delta $ given by
Eq.~(\ref{delta}), with $m^{2}=\mu _{i}^{2}+h_{i}^{2}\phi ^{2}$
 for each particle species $i$. Performing the momentum integration we obtain
\begin{equation}
\eta =\sum_{i}\frac{g_{i}h_{i}^{4}}{\Gamma }\int_{-\infty }^{+\infty
}c_{1i}^{2}\phi ^{2}\phi ^{\prime 2}dz.  \label{eta}
\end{equation}%
For $T\approx T_{c}$ we can use the thin wall
approximation\footnote{The wall is thick in comparison with
$T^{-1}$, but it is thin in comparison with the bubble radius.}
$\phi ^{\prime }= \sqrt{2V_T\left( \phi \right) }$ in Eq.
(\ref{eta}). Thus, the friction caused by particles with
thermal distributions is given by
\begin{equation}
\eta _{\mathrm{th}}=\sum_{i}\frac{g_{i}h_{i}^{4}}{\Gamma }\int_{0}^{\phi
_{c}}c_{1i}^{2}\left( \phi \right) \phi ^{2}\sqrt{2V_T}d\phi .  \label{eta2}
\end{equation}
Particles with larger couplings $h$ give the main contributions
to the friction force, since they have stronger interactions
with the bubble wall.

\subsection{Infrared bosons}

Infrared boson excitations must be treated classically
\cite{mt97} and undergo overdamped evolution \cite{asy97}.
Relating the population function to the squared amplitude of
the field \cite{m00}, one obtains the equation
\begin{equation}
\frac{\pi m_{D}^{2}}{8p}\frac{df}{dt}=-E^{2}\delta f,  \label{damp}
\end{equation}%
where $m_{D}^{2}$ is the squared Debye mass,
$m_{D}^{2}=(11/6)g^{2}T^{2}$ for the W and Z bosons of the SM,
and $m_{D}^{2}=h^{2}T^{2}/3$ for a scalar singlet. Writing
$f=f_{0}\left( E/T\right) +\delta f$ we have, to first-order in
$v_w$,
\begin{equation}
\delta f=-\frac{\pi m_{D}^{2}}{16pTE^{3}}f_{0}^{\prime }\frac{dm^{2}}{d\phi }
\phi ^{\prime }v_{w}. \label{depart}
\end{equation}%
Inserting the departure from equilibrium (\ref{depart}) in Eq.
(\ref{fric}) and doing the momentum integration, we obtain
\begin{equation}
\eta =\sum_{\mathrm{bosons}}g_{i}\frac{\pi m_{D}^{2}}{8T^{3}}
h^{4}\int_{-\infty }^{+\infty }b\phi ^{2}\phi ^{\prime 2}dz.  \label{etair}
\end{equation}%
with
\begin{equation}
b=-T^{2}\int \frac{d^{3}p}{\left( 2\pi \right) ^{3}pE^{4}}
f_{0}^{\prime }(E/T). \label{b}
\end{equation}%
The integral in Eq. (\ref{etair}) will diverge if  $m(\phi)$
vanishes. Indeed, for small $m/T$, the momentum integral
(\ref{b}) is infrared dominated. Therefore, we can make the
approximation $f_{0}^{\prime }\left( x\right) \simeq -1/x^{2}$,
and we obtain
\begin{equation}
b=T^{4}/\left( 8\pi ^{2}m^{4}\right) .  \label{bchco}
\end{equation}%
Hence, for $\mu =0$ the integral in Eq. (\ref{etair}) has a
logarithmic divergence at $\phi =0$. However, the kinetic
description (\ref{fric}) breaks down for very infrared
particles, which in the case $m\to 0$ dominate. The actual
contribution of  degrees of freedom  with $p\lesssim L_w^{-1}$
is  subdominant, because their wavelength cannot resolve the
thickness of the wall \cite{m00}. Thus, for small $\mu $ we
will cut off the integral at $m\left( \phi \right)
=L_{w}^{-1}$, and we have
\begin{equation}
\eta _{\mathrm{ir}}=\sum_{\mathrm{bosons}}g_{i}\frac{\pi m_{D}^{2}}{8T^{3}}
h^{4}\int_{\phi _{0}}^{\phi _{c}}d\phi \phi ^{2}\sqrt{2V_T}b\left( \phi
\right) ,  \label{etair2}
\end{equation}%
with $\phi _{0}=\sqrt{L_{w}^{-2}-\mu ^{2}}/h$ for $\mu
<L_{w}^{-1}$, and $ \phi _{0}=0$ for $\mu >L_{w}^{-1}$. For a
particle with small enough $h$, we will have $\phi _{0}>\phi
_{c}$. In such a case the particle will not contribute to the
friction and we shall set $\eta =0$.

\begin{figure}[hbt]
\centering \epsfxsize=10cm \leavevmode \epsfbox{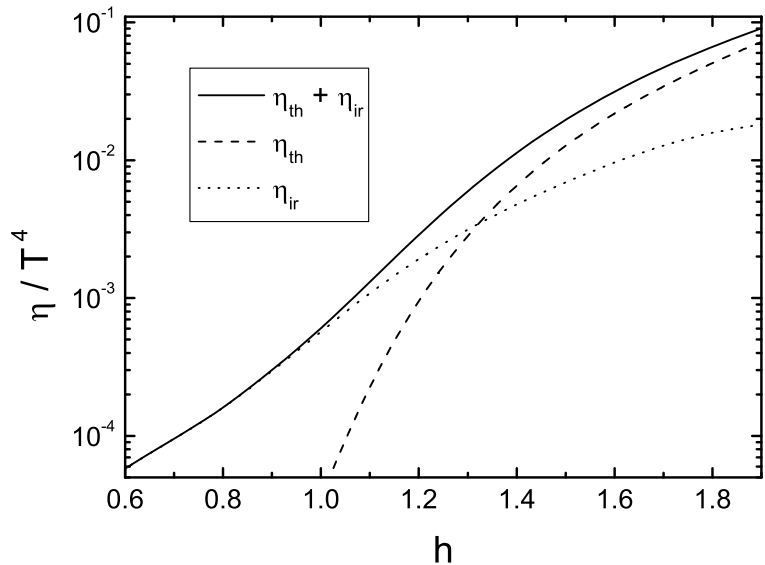}
\caption{The contribution of a complex singlet to the friction as a
function of the coupling $h$ for $\mu =0$ and $m_{H}=125GeV$.}
\label{etabos}
\end{figure}

\begin{figure}[hbt]
\centering \epsfxsize=10cm \leavevmode \epsfbox{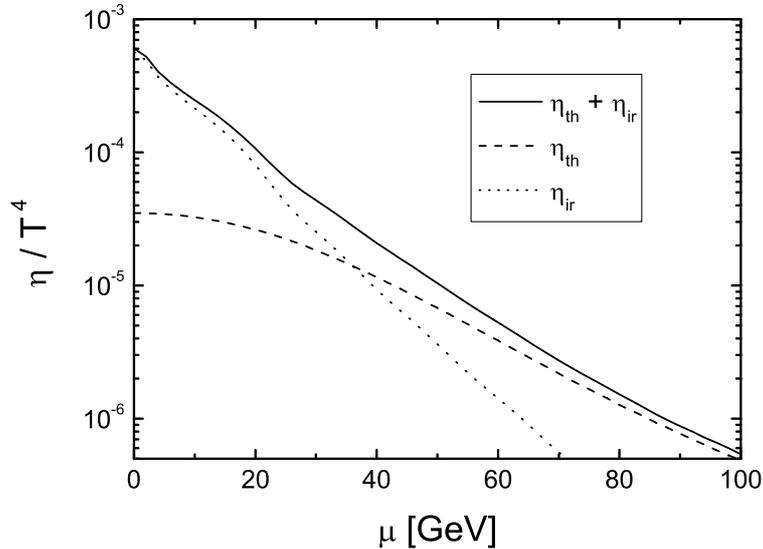}
\caption{The contribution of a complex singlet to the friction as a
function of $\mu$ for $h=1$ and $m_{H}=125GeV$.}
\label{etabosmu}
\end{figure}

Since Eqs. (\ref{eta2}) and (\ref{etair2}) depend on the
profile of the wall, the contribution of a particle species to
the friction depends on the whole particle content of the
model. As an example, we consider the case of the SM with an
additional complex scalar field.  Figure \ref{etabos} shows the
contribution of the extra singlet to the friction as a function
of the coupling $h$. The dashed line corresponds to the
friction coefficient $\eta_{\mathrm{th}}$ given by Eq.
(\ref{eta2}), the dotted line to $\eta_{\mathrm{ir}}$ given by
Eq. (\ref{etair2}), and the solid line shows their sum. In Fig.
\ref{etabosmu} we fix $h=1$ and vary $\mu$. The coefficients
$\eta_{\mathrm{th}}$ and $\eta_{\mathrm{ir}}$ dominate in
different parameter regions, and we shall use
$\eta=\eta_{\mathrm{th}}+\eta_{\mathrm{ir}}$ for applications
in this paper. As expected, $\eta_{\mathrm{ir}}$ dominates for
relatively small values of $\mu$ and $h$. We see that the
friction increases with $h$, as particles interact more
strongly with the wall, and decreases with $\mu$ due to
Boltzmann suppression of the particle density.

\subsection{Limiting cases}

In section \ref{modelos} we shall compute the friction
coefficient for each species numerically. It is useful, though,
to consider here some limiting cases.

The coefficient $ c_{1}$ depends on $m\left( \phi \right) $. It
is usual to consider only the lowest order in $m/T.$ In that
limit we have
\begin{equation}
c_{1f}=\log 2/2\pi ^{2},\quad c_{1b}=-\log \left( m/T\right) /2\pi ^{2},
\label{c1chco}
\end{equation}
for fermions and  bosons respectively. Since $c_{1b}$ depends
only logarithmically on $m$, we may replace $m\left( \phi
\right) \approx m\left( \phi _{c}\right) $ in Eq.
(\ref{c1chco}). The integral  in Eq. (\ref{eta2}) goes like
$\phi _{c}^{2}\sigma $, and in the small $m/T$ limit we recover
Eq. (\ref{etatth}),
\begin{equation}
\eta_{\mathrm{th}} \sim gh^{4}\frac{\phi _{c}^{2}\sigma }{\Gamma }\left(
\frac{\log \chi }{2\pi ^{2}}\right) ^{2}\quad \mathrm{for\ }m/T\ll 1,
 \label{etatth2}
\end{equation}%
with $\chi =2$ for fermions and $\chi =\sqrt{h^{2}\phi
_{c}^{2}+\mu ^{2}} /T\sim h$ for bosons. However, the
approximation (\ref{c1chco}) will break down whenever $\mu $ or
$h\phi $ become large in comparison with the temperature. In
particular, particles with large $h$ in a strong phase
transition will have $m\sim h\phi\gg T$  in the broken-symmetry
phase.

In the limit $m/T\gg 1$ we have
\begin{equation}
c_{1}=\left( m/T\right) ^{1/2}\exp \left( -m/T\right) /\left( 2\pi \right)
^{3/2}.  \label{c1gde}
\end{equation}%
It is apparent that the contribution of a heavy particle to the
friction is suppressed by a Boltzmann factor. In the case of
small $\mu $ but $h\phi _{c}\gg T$, we can estimate the
integral in Eq. (\ref{eta2}) using a quartic approximation for
the effective potential at $T=T_{c}$, $V_T\left( \phi \right)
\approx A\phi ^{2}\left( \phi _{c}-\phi \right) ^{2}$ with $
A=V_T^{\prime \prime }\left( 0\right) /(2\phi^{2} _{c})$. Since
for small $\phi $ the integrand is suppressed by a factor $\phi
^{3}$, we can use the approximation (\ref{c1gde}) in Eq.
(\ref{eta2}). To leading order in $h\phi _{c}/T_{c}$, we obtain
\begin{equation}
\eta_{\mathrm{th}} \approx \frac{3g}{32\pi^3}\frac{\sqrt{V_T^{\prime
\prime }\left(
0\right) }T_{c}^{4}}{\Gamma }\quad \mathrm{for\ }h\phi _{c}/T_{c}\gg 1.
\label{bounded}
\end{equation}%
We see that in the limit of a very strong phase transition the
friction coefficient does not increase like $h^{4}$ as in Eq.
(\ref{etatth2}). This is because the integrand in Eq.
(\ref{eta2}) is suppressed for $h\phi\gtrsim T$. Still, $\eta $
grows with the strength of the phase transition due to the
factor $\sqrt{ V_T^{\prime \prime }\left( 0\right) }$. For a
heavy particle with $\mu \gg T $, the function $c_{1}$ does not
depend on $\phi$, and the integral in Eq. (\ref{eta2}) goes
with $c_1^2 \phi_c^2\sigma$. Therefore, we have
\begin{equation}
\eta_{\mathrm{th}} \sim \frac{gh^{4}}{\left( 2\pi \right) ^{3}}\frac{\phi _{c}^{2}\sigma
}{\Gamma }{\frac{\mu }{T}}e^{-2\mu /T}\quad \mathrm{for\ }\mu /T\gg 1.
\end{equation}%
This exponential suppression of the contribution to the
friction is important, since the heavy particle will also be
decoupled from the effective potential. For instance, as the
value of $\mu $ for a boson is increased, the phase transition
becomes weaker. Consequently, there will be less supercooling
and a smaller pressure difference at $T=T_{n}$. Nevertheless,
the wall velocity will not necessarily decrease, since the
friction will also be smaller.

For the coefficient $b$, we have  the approximation
(\ref{bchco}) for  $m /T\ll 1$. For small $\phi $ we also have
$V_T\left( \phi \right) \approx \frac{1}{2}V_T^{\prime \prime
}\left( 0\right) \phi ^{2}$, and we obtain
\begin{equation}
\eta \approx \frac{gm_{D}^{2}\sqrt{V_T^{\prime \prime }
\left( 0\right)}T_{c}}{64\pi} \left[
\log \frac{m\left( \phi _{c}\right) }{m\left( \phi _{0}\right)}-
\frac{\left( m^{2}\left(
\phi _{c}\right) -m^2\left( \phi _{0}\right)\right)
\mu ^{2}}{2m^{2}\left( \phi _{c}\right)
m^2\left( \phi _{0}\right)}\right] \quad \mathrm{for\ } m/T\ll 1.
\end{equation}%
The factor of $h^{4}$ in (\ref{etair2}) has disappeared, but a
factor $h^{2}$ still  remains  in $m_{D}^2$. This is the
generalization of Eq. (\ref{etatir}) for small but nonvanishing
$ \mu $. Notice that for $\mu<L_w^{-1}$ we have
$m(\phi_0)=L_w^{-1}$, while for $\mu>L_w^{-1}$ we have
$m(\phi_0)=\mu$. Thus,  noting that $\sqrt{V_T^{\prime \prime
}\left( 0\right) }\sim L_{w}^{-1}$, we recover Eq.
(\ref{etatir}) in the limit $\mu \rightarrow 0$. On the other
hand, for $\mu\neq 0$ the friction is smaller.

In the limit of very large $m/T$, we have
\begin{equation}
b=e^{-m/T}T^{3}/\left( 2\pi ^{2}m^{3}\right) .
\end{equation}%
Let us consider, e.g., the case $\mu=0$ and $h\phi_c\gg T$.
Since $b$ is exponentially suppressed for $m=h\phi\gg T$, we
cut off the integral in Eq. (\ref{etair2}) at $\phi = T/h\ll
\phi _{c}$ and use the small $m$ approximation for $b$. Thus,
we obtain
\begin{equation}
\eta \sim \frac{gm_{D}^{2}\sqrt{V_T^{\prime \prime }
\left( 0\right)}T_{c}}{64\pi}
\log (TL_w)
 \quad \mathrm{for\ } h\phi_c/T_c \gg 1.
\end{equation}%
Due to the infrared behavior, the result is similar to the case
$\mu=0$, $h\phi_c/T_c\ll 1$. The only difference is that the
$\log$ is evaluated at $T$ instead of $m(\phi_c)$. For a
particle with large $\mu $ the contribution is
\begin{equation}
\eta \sim \frac{gh^{4}}{16\pi }
\frac{m_{D}^{2}\phi _{c}^{2}\sigma}{\mu ^{3}}
e^{-\mu /T}\quad \mathrm{for\ }\mu /T\gg 1,
\end{equation}%
which is exponentially suppressed, as expected.

\begin{figure}[hbt]
\centering \epsfxsize=10cm \leavevmode \epsfbox{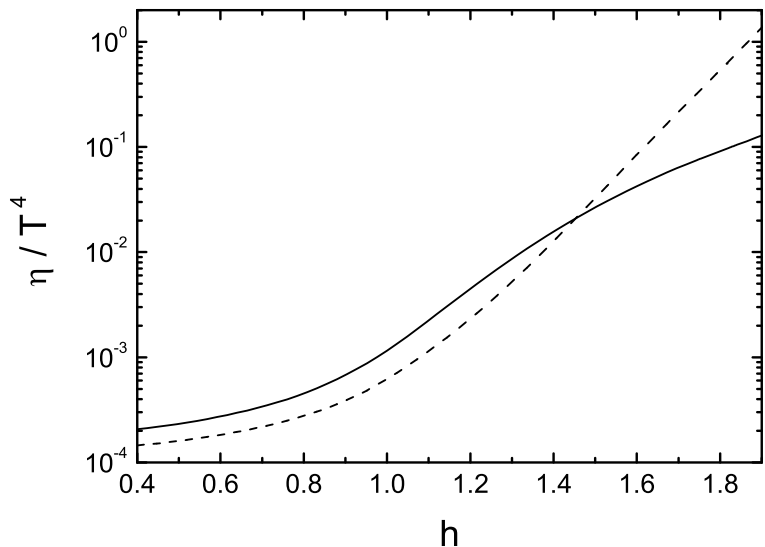}
\caption{The friction coefficient for the SM with a complex
singlet as a
function of the coupling $h$ for $m_{H}=125GeV$ and $\mu =0$
is plotted in solid line.
The dashed line corresponds to using the small $m/T$
approximations.}
\label{figfric}
\end{figure}

In Figure \ref{figfric} we have plotted the total friction
coefficient (including the contributions of the top quark and
gauge bosons) for the case of the SM with an additional
singlet, as a function of the coupling $h$ of the extra
singlet. The dashed line corresponds to using the
approximations given by Eqs. (\ref{etatth}) and (\ref{etatir}).
Due to numerical factors in these rough approximations, this
curve does not match the solid line for small $h$. Apart from
this fact, the behavior is similar. However, for larger vales
of $h$ the dashed line grows faster and crosses the solid line.
This is because the approximation increases with $ h^{4}$,
while the correct result does not, according to Eq.
(\ref{bounded}).

We wish to emphasize the fact that the friction depends on the
wall profile besides depending on the parameters $g$ and $h$.
For instance, in the case of small $ m/T$,  $\eta $ is
proportional to $\phi _{c}^{2}\sigma \sim \phi
_{c}^{4}L_{w}^{-1}$. Sometimes the wall width is estimated as $
L_{w}\approx 1/T$. However, $L_{w}$ may change significantly
with the strength of the phase transition. We have computed
$L_w$ and $\sigma$ numerically using Eqs. (\ref{sigma}) and
(\ref{lw}). For the parameters used in Fig. \ref{figfric} they
vary from $L_{w}\sim 100/T$ and $\sigma \sim 10^{-4}T^{3}$ for
small $h$ (weak phase transitions) to $L_{w}\sim 1/T$ and
$\sigma \sim T^{3}$ for large $h$ (strong phase transitions).
This dependence will become particularly important if the
strength of the phase transition is varied without changing the
particle content of the model (e.g., by changing the tree-level
potential).

\section{The bubble wall velocity} \label{veloc}

If we ignore hydrodynamics, the steady state velocity of the
bubble wall is obtained by equating the friction force
(\ref{fricnr}) to the pressure difference between phases. We
shall use subindexes $b$ and $u$ for quantities in the broken
and unbroken symmetry phase, respectively. Thus, ignoring
hydrodynamics, we have $\eta v_{w}=p_b(T)-p_u(T)$. Here, the
pressure difference  $\Delta p(T)=p_{b}\left( T\right)
-p_{u}\left( T\right)$ is given by the effective potential
(\ref{effpot}), $\Delta p(T) =-V\left( \phi _{m}\left( T\right)
,T\right) $. However, the motion of the wall in the plasma
causes variations of the temperature and velocity of the fluid,
and is in turn affected by these perturbations.

For hydrodynamic considerations (see, e.g.,
\cite{gkkm84,landau,h95}) we can assume a thin wall, and the
temperature and fluid velocity turn out to be discontinuous at
the interface. Thus, we have different temperatures $T_u$ and
$T_b$ on each side of the wall. We assume a planar wall in
stationary motion in the $z$ direction. In the rest frame of
the wall, the continuity conditions for energy and momentum
fluxes give the relations \cite{landau}
\begin{eqnarray}
w_{u}\gamma _{u}^{2}v_{u} &=&w_{b}\gamma _{b}^{2}v_{b},\   \label{eqlandau}
\\
w_{u}\gamma _{u}^{2}v_{u}^{2}+p_{u} &=&w_{b}\gamma _{b}^{2}v_{b}^{2}+p_{b},
\nonumber
\end{eqnarray}%
where $v$ is the fluid velocity $\gamma =1/\sqrt{1-v^{2}}$, and
$w=\rho+p$ is the enthalpy density. In Eq. (\ref{eqlandau}) we
have, e.g., $p_{u}\equiv p_{u}\left( T_{u}\right) $ (notice
that the thermodynamical quantities have in general different
values in each phase, even for $T_u=T_b$). To obtain a
macroscopic equation involving the friction, one can introduce
a damping term of the form $u^{\mu }\partial _{\mu }\phi $ in
the equation of motion for the Higgs field. This equation must
be integrated taking into account the temperature variation.
Assuming a thin wall, one obtains \cite{ms09}
\begin{equation}
p_{u}-p_{b}-\frac{1}{2}\left( s_{u}+s_{b}\right)
\left( T_{u}-T_{b}\right) +
\frac{\eta }{2}\left( v_{u}\gamma _{u}+v_{b}\gamma _{b}\right)
=0,
\label{eqmicro}
\end{equation}%
where $s$ is the entropy density and $\eta $ is the friction
coefficient obtained from the microphysics calculation, $\eta
=\left[\Delta p(T)/v_{w}\right]_{ \mathrm{micro}}$. The
thermodynamical variables are related by an equation of state
(EOS), so Eqs. (\ref{eqlandau}) and (\ref{eqmicro}) have only
four unknowns, namely, the velocities $v_{u,b}$ and the
temperatures $ T_{u,b}$. Besides, the temperature $T_{u}$
outside the bubble can be determined by computing the
nucleation temperature.

The velocities $v_{u}$ and $v_{b}$ of the fluid are measured in
the reference frame of the wall. Equivalently, $\left\vert
v_{u}\right\vert$ and $\left\vert v_{b}\right\vert$ give the
velocity of the wall with respect to the fluid on each side of
the wall. We want to calculate the wall velocity $v_w$ in the
reference frame in which the fluid is at rest \emph{very far}
in front of the wall and very far behind the wall (at the
center of the bubble). We shall refer to this reference frame
as the laboratory frame.

The stationary motion of the wall admits two kinds of
solutions, called detonations and deflagrations. For
detonations we have $\left\vert v_{b}\right\vert <\left\vert
v_{u}\right\vert $ and $\left\vert v_{u}\right\vert
>c_{s}$, where $c_{s}=1/\sqrt{3}$ is the speed of sound in the
relativistic fluid. Therefore, in this case the wall moves
supersonically with respect to the fluid in front of it. Hence,
no information on the motion of the wall is transmitted into
this fluid, which can be assumed to be at rest in the
laboratory frame. According to this boundary condition, the
detonation is supersonic, $ v_{w}=-v_{u}>c_s$, and the
temperature $T_{u}$ is the nucleation temperature $T_{n}$.  In
the reference frame of the wall, the incoming flow is
supersonic. The outgoing flow inside the bubble has a lower
velocity and could in principle be subsonic (strong
detonation). This possibility, however, is forbidden by the
boundary condition at the center of the bubble. Therefore a
detonation can only be weak ($\left\vert v_{b}\right\vert
>c_{s}$) or Jouguet ($\left\vert v_{b}\right\vert =c_{s}$). In
general, though, we will only have weak detonations
\cite{ms09}.

For deflagrations we have $\left\vert v_{b}\right\vert
>\left\vert v_{u}\right\vert $ and $\left\vert v_{u}\right\vert
<c_{s}$, so the wall is subsonic with respect to the fluid in
front of it. It turns out that a single front is not enough to
satisfy the boundary conditions in this case. Thus, the
phase-transition front must be preceded by a shock front. In
the laboratory frame, the fluid between the two fronts has a
finite velocity, and outside this region the fluid is at rest.
Therefore, we have $v_{w}=-v_{b}$, and the fluid in front of
the wall moves in the direction of the latter. In the frame of
the wall, there is a subsonic flow coming from the symmetric
phase, and an outgoing flow with a larger velocity $v_{w}$. In
principle, the outgoing flow can be supersonic (strong
deflagration), although in general we will have $ v_{w}<c_{s}$
(weak deflagration). The case $v_w=c_s$ is called a Jouguet
deflagration. In the deflagration case, the temperature $
T_{u}$ is higher than $T_{n}$, since the fluid in the region
between the bubble wall and the shock front is compressed and
reheated. The relation between $T_{u}$ and $T_{n}$ can be
obtained by considering the fluid conditions (\ref{eqlandau})
for the shock front \cite{landau}.

In order to solve Eqs. (\ref{eqlandau}) and (\ref{eqmicro}), it
is convenient to use the bag equation of state
\begin{eqnarray}
\rho _{u}\left( T\right) =a_{u}T^{4}+\varepsilon , &&p_{u}\left( T\right)
=a_{u}T^{4}/3-\varepsilon ,  \label{eos} \\
\rho _{b}\left( T\right) =a_{b}T^{4}, &&p_{b}\left( T\right) =a_{b}T^{4}/3,
\nonumber
\end{eqnarray}
which is the simplest EOS that keeps the essential features of
a phase transition. The coefficients $a_{u}$ and $a_{b}$ must
be different, since in this model we have
$a_{b}/a_{u}=1-3\alpha _{c}$, with $\alpha _{c}\equiv
\varepsilon /\left( a_{u}T_{c}^{4}\right) $. The constant
$\varepsilon $ is the false vacuum energy density of the
high-temperature phase, and also determines the latent heat
$L\equiv \rho _{u}\left( T_{c}\right) -\rho _{b}\left(
T_{c}\right) $ by\footnote{A simpler model with a single
coefficient $a_{u}=a_{b}=a$ is sometimes used. In that case we
would have $L=\varepsilon $. However, such a model is not
suitable for a phase transition, since the critical temperature
[for which $ p_{+}(T_c)=p_{-}(T_c)$] does not exist
\cite{ms09}.}
\begin{equation}
L=4\varepsilon .  \label{epslat}
\end{equation}%
In the general case, the latent heat and the vacuum energy
density will not have this simple relation. In order to apply
the results of this calculation to a general model, it is
convenient to rewrite the parameter $\alpha _{c}$ as
\begin{equation}
\alpha _{c}=\frac{L}{4\tilde{\rho}_{u}(T_{c})}, \label{ac}
\end{equation}%
where $\tilde{\rho}_{u}\left( T\right) =a_{u}T^{4}$ is the
\emph{thermal} energy density in the high-temperature phase.

Using the EOS (\ref{eos}) in Eqs. (\ref{eqlandau}) and
(\ref{eqmicro}), we eliminate $T_{b},$
\begin{equation}
\frac{T_{b}}{T_{u}}=\left[ \frac{a_{u}}{a_{b}}\left( 1-\alpha _{u} \frac{
1+v_{u}v_{b}}{1/3-v_{u}v_{b}}\right) \right] ^{1/4},
\end{equation}%
where $\alpha _{u}\equiv \varepsilon /\left(
a_{u}T_{u}^{4}\right) $, and we still have two equations for
$v_{u}$, $v_{b}$ and $\alpha _{u}$. One of them comes from
hydrodynamics alone \cite{s82},
\begin{equation}
v_{u}=\frac{\frac{1}{6v_{b}}+\frac{v_{b}}{2}\pm \sqrt{\left( \frac{1}{6v_{b}}
+\frac{v_{b}}{2}\right) ^{2}+\alpha _{u}^{2}+\frac{2}{3}\alpha _{u}- \frac{1
}{3}}}{1+\alpha _{u}},  \label{steinhardt}
\end{equation}%
where the + and - signs in front of the square root correspond
to detonations and deflagrations, respectively. The second
equation involves microphysics \cite{ms09},
\begin{equation}
\frac{4v_{u}v_{b}\alpha _{u}}{1-3v_{u}v_{b}}-\frac{2}{3}\left( 1+ \frac{s_{b}
}{s_{u}}\right) \left( 1-\frac{T_{b}}{T_{u}}\right) +\frac{2\alpha _{u}\eta}{
L}\left( \left\vert v_{u}\right\vert \gamma _{u}+\left\vert v_{b}\right\vert
\gamma _{b}\right) =0,  \label{eqfric}
\end{equation}%
with $s_{b}/s_{u}=\left( a_{b}/a_{u}\right) \left(
T_{b}/T_{u}\right) ^{3}$. In the case of detonations, we can
solve these equations for $\left\vert v_{u}\right\vert =v_{w}$
as a function of $\alpha _{u}$, since $\alpha _{u}$ depends on
$T_{u}=T_{n}$. In the case of deflagrations, we have an
additional equation relating $T_{u}$ and $T_{n},$
\begin{equation}
\frac{\sqrt{3}\left( \alpha _{n}-\alpha _{u}\right) }{\sqrt{\left( 3\alpha
_{n}+\alpha _{u}\right) \left( 3\alpha _{u}+\alpha _{n}\right) }}= \frac{
v_{u}-v_{b}}{1-v_{u}v_{b}},
\end{equation}%
where $\alpha _{n}\equiv \varepsilon /\left(
a_{u}T_{n}^{4}\right) $, and we can solve for $\left\vert
v_{b}\right\vert =v_{w}$. In any case, $T _{u}$ can be
eliminated, and the result depends on
\begin{equation}
\alpha _{n}=\frac{L}{4\tilde{\rho}_{u}(T_{n})}.  \label{an}
\end{equation}

These equations can be solved numerically \cite{ms09}, and the
wall velocity finally depends on the parameters $\alpha _{c}$
and $\alpha _{n}$, and on the ratio $\eta /L$. Depending on the
parameters, there may be only deflagrations, only detonations,
both, or none. The deflagration solution always exists if the
friction is large enough \emph{or} the supercooling is small
enough. The detonation solution will exist if the friction is
small enough \emph{and} the supercooling is large enough. In
general, there is no detonation solution when the deflagration
velocity is smaller than $\approx 0.5$. As the velocity
approaches the speed of sound, the detonation solution may
appear. The deflagration solution disappears soon after
becoming supersonic. Analytical approximations were also found
in Ref. \cite{ms09} for both solutions. In this paper we shall
use the numerical results.

Notice that the definitions of $\alpha _{c}$ and $\alpha _{n}$
[Eqs. (\ref{ac}) and (\ref{an})] involve the thermal energy
densities at the temperatures $T_c$ and $T_n$, but the latent
heat is in both cases the energy density discontinuity at
$T=T_c$. These parameters arise as a consequence of the use of
the simple bag EOS, and their definitions must be respected in
applications. As discussed in Ref. \cite{ms09}, if we used for
$\alpha _{n}$ the energy density that is released at $T=T_{n}$,
which is larger than $L$, we would be overestimating the
velocity, since this would be equivalent to considering a
stronger supercooling [i.e., a smaller value of
$\tilde{\rho}_{u}(T_{n})$].

\section{The electroweak wall velocity} \label{modelos}

In this section we calculate the value of the wall velocity in
the electroweak phase transition. For that aim, we compute the
parameters $\alpha _{c}$, $\alpha _{n}$, and $\eta /L$ for
several models. The temperature $T_n$ is defined through Eq.
(\ref{intnucl}), and is the temperature at which bubbles begin
to nucleate. By the time bubbles occupy all space, the
temperature will be in general different. If the wall velocity
is relatively large, bubbles will not have time to interact
with each other until they percolate. In this case the
temperature will just decrease due to the expansion of the
Universe, and the velocity will increase during the phase
transition. In general, though, the transition will be short
enough, so that the velocity will not change significantly from
$v_{w}\left( T_{n}\right)$ \cite{mege00}. On the other hand, in
the case of slow deflagrations, the supersonic shock fronts
preceding the walls will cause a reheating {during} bubble
expansion. Hence, the temperature will grow and the wall
velocity will decrease from $v_{w}\left( T_{n}\right)$.
Depending on the amount of latent heat, the reheating
temperature  may be very close to $T_{c}$ \cite{m04,ms08,h95}.
In this case, the wall velocity will decrease significantly. We
shall not take into account this possibility in the present
paper. We remark, though, that it may have important
consequences \cite{h95,m01}.

We shall consider several extensions of the SM. The relevant SM
contributions to the one-loop effective potential come from the
$Z$ and $W$ bosons, the top quark, and the Higgs and Goldstone
bosons.
It is usual to ignore the Higgs sector in the one-loop
radiative corrections.
This should be a good approximation in extensions of the SM
which include particles with strong couplings to $\phi $. The
$\phi $-dependent masses of the weak gauge bosons and top quark
are of the form $ h_{i}\phi $, with $h_{i}=m_{i}/v$, where
$m_{i}$ are the physical masses at zero temperature. We shall
ignore, as usual, the longitudinal components of the weak gauge
bosons, which are screened by plasma effects. Thus, the $W$ and
$Z$ contribute corrections of the form
(\ref{v1loop},\ref{f1loop}) to the free energy (\ref{ftot}),
with $4$ and $2$ bosonic d.o.f., respectively. The top
contributes with $g_{t}=12$ fermionic d.o.f. The rest of the SM
particles have $h_{i}\ll 1$ and only contribute a $\phi
$-independent term $-\pi ^{2}g_{\mathrm{light}}T^{4}/90$, with
$g_{\mathrm{light}}\approx 90$. To strengthen the electroweak
phase transition, extra particles are usually added to the SM,
with strong couplings to $\phi $.

\subsection{SM extensions with scalars}

It is well known that the easiest way of strengthening the phase transition
is by extending the scalar sector of the SM. The simplest extension consists
of adding gauge singlet scalars \cite{ah92,dhss91}.

\subsubsection{A complex singlet}

We shall first consider a model in which a single complex
scalar $S$ is added to the SM \cite{cv93,eq93,a07}. The
coupling to the Higgs is of the form $2h^{2}S^{\dag }SH^{\dag
}H$, and the field $S$ may have a $SU(2)\times U(1)$ invariant
mass term $\mu ^{2}S^{\dag }S$ and a quartic term $\lambda
_{S}\left( S^{\dag }S\right) ^{2}$. For simplicity, we will
ignore the possibility that cubic terms exist in the tree-level
potential, and we will assume $\mu ^{2}\geq 0$. We notice
however that a negative value of $\mu ^{2}$, as well as a cubic
coupling, may enhance the strength of the phase transition
through tree-level effects \cite{cv93}. We shall study the
effect of a cubic term at the end of this section. Hence, the
field $S$ gives contributions to the free energy of the form
(\ref{v1loop},\ref{f1loop}), with $g=2$ d.o.f. and a mass
$m^{2}\left( \phi \right) =h^{2}\phi ^{2}+\mu ^{2}$. The
thermal mass is given by $\Pi =\left( h^{2}+\lambda _{S}\right)
T^{2}/3$ \cite{eq93}. We shall take $ \lambda _{S}=0$ for our
numerical calculations. We have checked that considering
$\lambda _{S}\neq 0$ does not introduce qualitative differences
in the results.

\begin{figure}[hbt]
\centering \epsfxsize=10cm \leavevmode \epsfbox{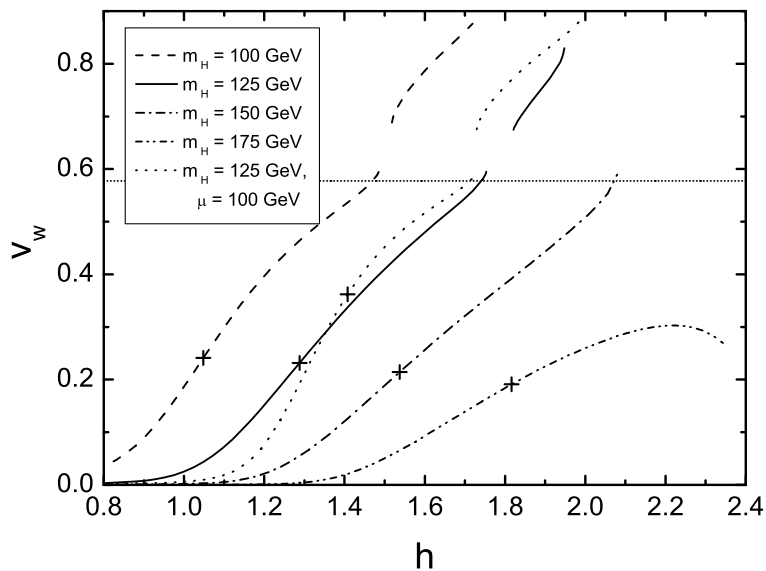}
\caption{The wall velocity as a function of the coupling $h$ of the singlet,
for several Higgs masses. All the curves correspond to the case $\protect\mu
=0$, except for the doted line, which corresponds to $\protect\mu=100GeV$.
The crosses indicate the case $\protect\phi_n/T_n=1$.}
\label{figsing}
\end{figure}

Figure \ref{figsing} shows the value of the wall velocity at
the nucleation temperature $T_{n}$, as a function of the
coupling $h$ for different values of the parameters. We have
plotted both the deflagration and the detonation solutions,
when they exist, for the case $\mu =0$ (which maximizes the
strength of the phase transition) and Higgs masses
$m_{H}=100,125,150$ and $ 175GeV$, and for the case $\mu
=100GeV$ and $m_{H}=125GeV$. For the same $h$, larger values of
$m_{H}$ give weaker phase transitions and, consequently, lower
values of $v_{w}$. The crosses indicate the points where $\phi
\left( T_{n}\right) /T_{n}=1$. As $h$ is increased, the phase
transition becomes stronger. As a consequence, the temperature
$T_{n}$ decreases and the supercooling stage lasts longer. In
this model, if $h$ is too large the Higgs VEV remains stuck in
the symmetric phase, and the Universe enters a period of
inflation. We are not interested in such extreme cases, and we
just plot the curves up to a value of $h$ for which the time
required to get out of the supercooling stage becomes too long
for the numerical computation.

Although the amount of supercooling grows with $h$, so does the
friction and the behavior of the wall velocity is rather
unpredictable. As can be seen in Fig. \ref{figsing}, the
velocity in general increases with $h$. However, for larger
values of $m_{H}$ (e.g., the case $m_H=175GeV$), the velocity
eventually decreases for large $h$. This is because larger
values of $h$ are required to achieve a phase transition of the
same strength, and the friction becomes important. Only the
cases $m_{H}=100$ and $125GeV$ have detonations (the short
curves which lie above the speed of sound). Notice that there
is a small range of $h$ for which there are neither
deflagrations nor detonations, i.e., no stationary solution for
the wall velocity. In such a case, the steady state will not be
reached, and the wall will accelerate until the end of the
phase transition. The case in which the wall velocity gets
close to the speed of light may have important implications for
gravitational wave generation. Such ultrarelativistic
velocities have been considered recently \cite{bm09} for the SM
extension with singlet scalar fields, finding that ``runaway''
solutions exist for very strong phase transitions.

\begin{figure}[hbt]
\centering \epsfxsize=10cm \leavevmode \epsfbox{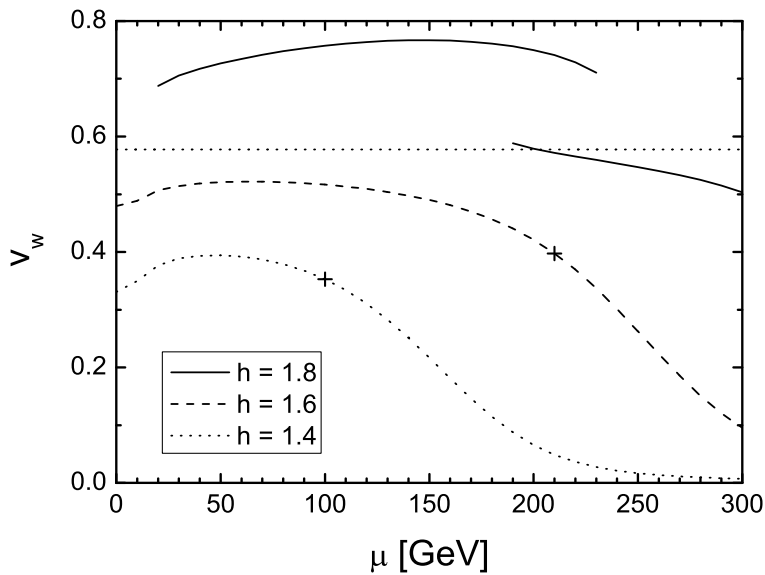}
\caption{The wall velocity as a function of $\protect\mu$, for $m_{H}=125GeV$
and three values of $h$.}
\label{figsingmu}
\end{figure}

For $\mu \neq 0$ the phase transition is weaker and the amount
of supercooling is lower, but the friction is lower too. As an
example, consider the case $m_{H}=125GeV$ and $\mu =100GeV$
(dotted line in Fig. \ref{figsing}). We see that for small $h$
the velocity is lower than in the case of $\mu =0$, while for
large $h$ the velocity is higher. In Fig. \ref{figsingmu} we
have plotted the wall velocity as a function of $\mu $ for
$m_{H}=125GeV$ and $h=1.4,$ $1.6$ and $1.8$. We see that for
small $\mu $ the wall velocity increases with $\mu $,
indicating that the friction decreases faster than the strength
of the phase transition. For large $\mu $ the velocity
eventually decreases, as the extra boson decouples from the
thermal plasma. For the cases $h=1.4$ and $h=1.6$ there are
only deflagrations. The crosses in the curves indicate a phase
transition with $\phi \left( T_{n}\right) /T_{n}=1$. To the
right of this point the phase transition is weaker. The value
$h=1.8$ is quite large, and there are detonations for a long
range of values of $\mu $. In this case, deflagrations appear
only for $\mu \gtrsim 170GeV$. We see that there is a range of
parameters for which there exist both kinds of solution for the
wall propagation, and also a range in which there is no
solution.

\subsubsection{A hidden sector}

Recently, an extension of the SM with several real singlets
$S_{i}$ has been considered \cite{eq07}. These bosons
constitute a hidden sector which couples only to the SM Higgs
doublet through a term $h^{2}H^{\dag }H\sum S_{i}^{2}$ (for
simplicity, universal couplings $h_{i}=h$ are assumed).
Following Ref. \cite{eq07}, we assume there are no linear,
cubic, or quartic terms in the hidden-sector scalar fields.
Therefore, this case is similar to the previous one, with
$m^{2}\left( \phi \right) =h^{2}\phi ^{2}+\mu ^{2}$ and $g$
d.o.f., where $g$ is the number of singlets.  We shall take
$g=12$, as in Ref. \cite{eq07}. If the fields $S_{i}$ do not
have mass terms, so that they only get a mass from electroweak
breaking, the phase transition can be made exceedingly strong.
Interestingly, this model allows to consider the classically
conformal case, corresponding to $m^{2}=0$ in Eq. (\ref{v0}).
The loop corrections break conformal invariance, and a mass
scale appears via dimensional transmutation. Imposing
appropriate renormalization conditions, for a given Higgs mass
the classically conformal case occurs for a fixed value of $h$.
Since we are interested in the variation of the strength of the
phase transition as a function of the parameters, we shall not
give special attention to this particular case.

\begin{figure}[hbt]
\centering \epsfxsize=10cm \leavevmode \epsfbox{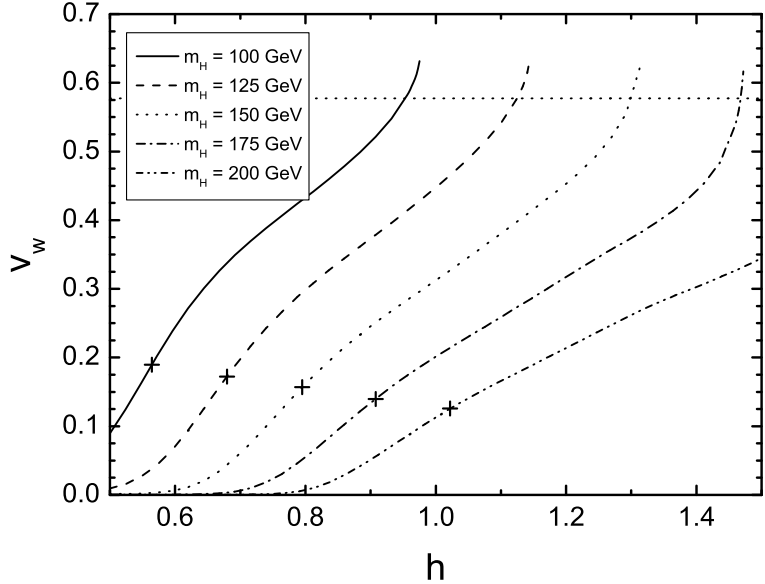}
\caption{The wall velocity for $g=12$ real scalars with $\protect\mu=0.$}
\label{fighid}
\end{figure}

Figure \ref{fighid} shows the velocity $v_{w}\left(
T_{n}\right) $ as a function of the coupling $h$ for $\mu =0$
and different values of $m_{H}$ in the range $100-200GeV$.
Since this model has more d.o.f. than the previous one, the
phase transition becomes strongly first-order for lower values
of $h $. As a consequence, the wall velocity grows more quickly
with $h$. However, we see that in all the cases the nucleation
stage becomes too long before detonations can exist. For $\mu
\neq 0$ the phase transition weakens and higher values of $h$
can be considered. We have found that detonations appear in the
range $\mu \sim 50-150GeV$ for low values of $m_{H}$ and
extreme values of $h$. For most values of $h$ there are only
deflagrations, and the behavior with $\mu $ is similar to that
of the lower curves of Fig. \ref{figsingmu}.

\subsection{The MSSM}

The Minimal Supersymmetric Standard Model (MSSM) has been
extensively investigated in connection to electroweak
baryogenesis \cite{ewbmssm}, since it may provide a strong
phase transition and sources of $CP$ violation. The MSSM
contains two complex Higgs doublets $H_{1}$ and $H_{2}$. We
define the vacuum expectation values $v_{1}\equiv \langle
H_{1}^{0}\rangle $ and $v_{2}\equiv \langle H_{2}^{0}\rangle $.
It is customary to simplify the problem by considering the
limit in which the CP-odd Higgs mass is large ($m_{A}\gg
m_{Z}$). In this limit the low energy theory contains a single
Higgs doublet $\Phi $, and the masses and couplings depend on
$\tan \beta \equiv v_{2}/v_{1}$. Thus, calling $\phi /\sqrt{2}$
the background of the real neutral component of $\Phi $, the
tree-level potential is of the form (\ref{v0}), with the
quartic coupling given by $\lambda =\left( g^{2}+g^{\prime
2}\right) \cos ^{2}\left( 2\beta \right) /8$. Therefore, the
tree-level Higgs mass is bounded by $m_{H}^{2}<m_{Z}^{2}$.
However, this tree-level relation is spoiled by radiative
corrections (see e.g. \cite{ceqr95}) and we shall consider
$m_{H}$ as a free parameter.

In this model, the relevant SM field-dependent masses are those
of the gauge bosons,
\begin{equation}
m_{W}^{2}=g^{2}\phi ^{2}/4\equiv h_{W}^{2}\phi ^{2},\quad m_{Z}^{2}=\left(
g^{2}+g^{\prime 2}\right) \phi ^{2}/4\equiv h_{Z}^{2}\phi ^{2},
\end{equation}%
and  top quark,
\begin{equation}
m_{t}^{2}\left( \phi \right) =\frac{h_{t}^{2}\sin ^{2}\beta }{2}\phi
^{2}\equiv \bar{h}_{t}^{2}\phi ^{2},
\end{equation}%
where $h_{t}$ is the Yukawa coupling to $H_{2}^{0}$. We shall work in the
limit in which the left handed stop is heavy ($m_{Q}\gtrsim 500GeV$). In
this case, the one-loop correction to the SM is dominated by the
right-handed top squark contribution, with the field-dependent mass given by
\begin{equation}
m_{\tilde{t}}^{2}\left( \phi \right) \approx m_{U}^{2}+h_{\tilde{t}}^{2}\phi
^{2},
\end{equation}%
where
\begin{equation}
h_{\tilde{t}}^{2}=0.15h_{Z}^{2}\cos 2\beta +\bar{h}_{t}^{2}\left( 1-\frac{
\tilde{A}_{t}^{2}}{m_{Q}^{2}}\right) ,
\end{equation}%
$m_{U}^{2}$ and $m_{Q}^{2}$ are soft breaking parameters, and
$\tilde{A}_{t}$ is the stop mixing parameter. If the mass of
the right-handed stop is of the order of the top mass or below,
the one-loop effective potential (\ref{effpot}) admits the
high-temperature expansion \cite{cqw98}
\begin{equation}
V\left( \phi ,T\right) =D\left( T^{2}-T_{0}^{2}\right) \phi ^{2}-T\left(
E_{SM}\phi ^{3}+6\frac{\mathcal{M}_{\tilde{t}}\left( \phi \right) ^{3}}{
12\pi } \right) +\frac{\lambda \left( T\right) }{4}\phi ^{4},  \label{fmssm}
\end{equation}%
where $D=m_{H}^{2}/\left( 8v^{2}\right)
+5h_{W}^{2}/12+5h_{Z}^{2}/24+h_{t}^{2}/2$ \cite{amnr02},
$T_{0}^{2}=m_{H}^{2}/(4D)$, $E_{SM}$ is the cubic-term
coefficient in the high-temperature expansion for the SM
effective potential, $E_{SM}\approx \left(
2h_{w}^{3}+h_{z}^{3}\right) /6\pi $, and
$\mathcal{M}_{\tilde{t}}^{2}\left( \phi \right)
=m_{\tilde{t}}^{2}\left( \phi \right) +\Pi _{\tilde{t}}\left(
T\right) $. The thermal mass is given by \cite{cqw98}
\begin{equation}
\Pi _{\tilde{t}}\left( T\right) =\left[ \frac{4g_{s}^{2}}{9}+ \frac{h_{t}^{2}
}{6}\left( 1+\sin ^{2}\beta \left( 1-\frac{\tilde{A}_{t}^{2}}{m_{Q}^{2}}
\right) \right) +\left( \frac{1}{3}-\frac{\left\vert \cos 2\beta \right\vert
}{18}\right) g^{\prime 2}\right] T^{2},
\end{equation}%
where $g_{s}$ is the strong gauge coupling. Following Ref. \cite{amnr02}, we
shall set $\tilde{A}_{t}=0$ for simplicity in the numerical calculation. We
shall also take $\sin ^{2}\beta =0.8$.

The phase transition strength is maximized for negative values
of the soft mass squared $m_{U}^{2}\approx -\Pi
_{\tilde{t}}\left( T\right) $ \cite{cqw96}, for which the
contribution of the term $\mathcal{M}_{\tilde{t}}^{3}$ in
(\ref{fmssm}) is of the form $-E_{MSSM}T\phi ^{3}$, with a
coefficient $ E_{MSSM}$ that may be one order of magnitude
larger than that of the SM. This would make the phase
transition sufficiently strong for baryogenesis for Higgs
masses as large as $100GeV$. However, such large negative
values of $m_{U}^{2}$ may induce the presence of color breaking
minima at zero or finite temperature \cite{cw95}. Demanding the
absence of such dangerous minima constrains the Higgs mass to
unrealistic values. Nevertheless, the two-loop corrections are
very important and can make the phase transition strongly
first-order even for $m_{U}\approx 0$ \cite{e96}. The most
important two-loop corrections are of the form $\phi ^{2}\log
\phi $ and are induced by the SM weak gauge bosons, as well as
by stop and gluon loops \cite{e96,bd93}. In the case of a heavy
left-handed stop we have \cite{cqw98}
\begin{eqnarray}
V_{2}\left( \phi ,T\right) &\approx &\frac{\phi ^{2}T^{2}}{32\pi ^{2}}\left[
\frac{51}{16}g^{2}-3\left( 2\bar{h}_{t}^{2}\left( 1- \frac{\tilde{A}_{t}^{2}
}{m_{Q}^{2}}\right) \right) ^{2}\right.  \label{twoloop} \\
&&\left. +8g_{s}^{2}2\bar{h}_{t}^{2}\left( 1-\frac{\tilde{A}_{t}^{2}}{
m_{Q}^{2}}\right) \right] \log \left( \frac{\Lambda _{H}}{\phi }\right) ,
\nonumber
\end{eqnarray}%
where the scale $\Lambda _{H}$ depends on the finite corrections and
is of order $100GeV$. Following \cite{amnr02}, we will set $\Lambda
_{H}=100GeV$ for the numerical computation, given the slight
logarithmic dependence of $ V_{2}$ on $\Lambda _{H}$.

The phase transition takes place at some temperature between
the critical temperature $T_{c}$ and the temperature at which
the barrier between minima disappears. The latter is
approximately given by the parameter $T_{0}$ in Eq.
(\ref{fmssm}). In order to avoid the presence of color-breaking
minima, we only consider values of $m_{U}^{2}$ for which
$m_{U}^{2}+\Pi _{\tilde{t}}\left( T_{0}\right)
>0$ \cite{amnr02}.
For the computation of the temperature $T_{n}$ we used the
high-temperature approximation (\ref{fmssm}) for the one-loop
effective potential, together with the two-loop correction
(\ref{twoloop}). Therefore, for the friction coefficient we
also used the high-temperature approximation, given by Eqs.
(\ref{etatth}) and (\ref{etatir}). For the stop contribution we
have $m_{D}\sim m_{\tilde{t}}\sim h_{\tilde{t}}T$.

\begin{figure}[hbt]
\centering \epsfxsize=10cm \leavevmode \epsfbox{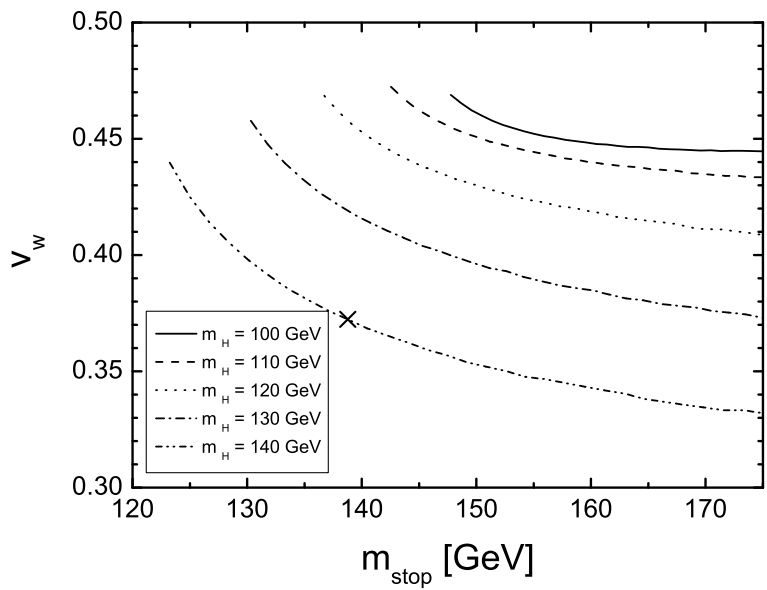}
\caption{The wall velocity as a function of the stop mass for several Higgs
masses. The cross in the lower curve marks
the point of $\protect\phi_n/T_n=1$.}
\label{figmssm}
\end{figure}

Figure \ref{figmssm} shows the wall velocity as a function of
the stop mass for Higgs masses $m_{H}=100,110,120,130$ and
$140GeV$. In the lower curve ($m_H=140GeV$), the cross marks
the case $\phi_n/T_n=1$. To the right of the cross the phase
transition is weak. As can be seen, for this model the velocity
is not very sensitive to the stop and Higgs masses. We only
obtain deflagrations, with velocities in the range $v_{w}\sim
0.3-0.45$.

\subsection{TeV fermions}

So far we have considered extensions of the SM for which the
relevant contributions to the effective potential came from new
scalars. In Ref. \cite{cmqw05}, it was shown that in extensions
with extra fermions strongly coupled to the Higgs field, the
phase transition may be sufficiently strong to avoid erasure of
the baryon asymmetry in the broken-symmetry phase. Strongly
coupled fermions, however, make the vacuum unstable unless the
Higgs is heavy. In the model considered in Ref. \cite{cmqw05},
this problem was solved by adding heavy bosons with similar
couplings and number of degrees of freedom, but with a large
$\phi $-independent mass term, so that they are decoupled from
the dynamics at $T\sim v$. The model can be considered as a
particular realization of split supersymmetry, where the
standard relations between the Yukawa and gauge couplings are
not fulfilled. Therefore, the fermions are higgsinos and
gauginos, with a total of 16 d.o.f.

Depending on the values of the Yukawa couplings and of the mass
parameters, the mass eigenvalues can be rather cumbersome. In the
simplest case, only $g=12$ d.o.f. are coupled to the SM Higgs, with
degenerate eigenvalues of the form $m_{f}^{2}\left( \phi \right) =\mu
^{2}+h^{2}\phi ^{2}$. One can assume for simplicity that the bosonic
stabilizing fields have the same number of d.o.f., and a dispersion
relation $m_{S}^{2}\left( \phi \right) =\mu _{S}^{2}+h^{2}\phi ^{2}.$
We shall also assume, as in Ref. \cite{cmqw05}, that $\Pi _{S}=0$.
The maximum value of $\mu _{S}$ consistent with stability is obtained
by requiring the quartic term in Eq. (\ref{v1loop}) to be positive at
scales much larger than $v.$ Taking into account only the radiative
corrections associated with the strongly coupled fields, one finds
that $\mu _{S}^{2}$ must be below the value
\begin{equation}
\mu _{S}^{2}=\exp \left( \frac{m_{H}^{2}8\pi ^{2}}{gh^{4}v^{2}}\right)
m_{f}^{2}\left( v\right) -h^{2}v^{2}.  \label{mus}
\end{equation}%
In order to minimize the effect of the stabilizing bosons on
the strength of the phase transition, we will set $\mu _{S}$ to
this maximum value. Notice, however, that we have $\mu _{S}\gg
\mu $ only for small $h$, so in general the stabilizing bosons
are not completely decoupled from the thermal plasma.

\begin{figure}[hbt]
\centering \epsfxsize=10cm \leavevmode \epsfbox{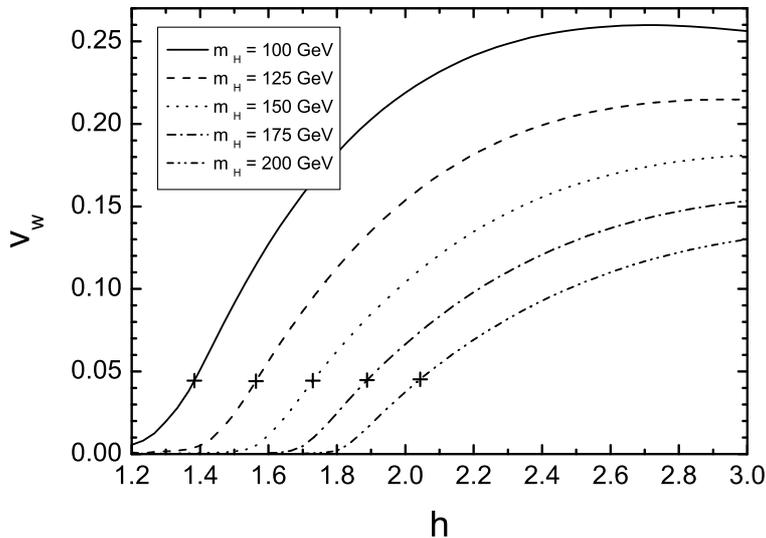}
\caption{The wall velocity as a function of $h$, for $g=12$  and $
\protect\mu=0$. The points to the right of the crosses correspond to phase
transitions with $\protect\phi_n/T_n>1$.}
\label{figfer}
\end{figure}

\begin{figure}[hbt]
\centering \epsfxsize=10cm \leavevmode \epsfbox{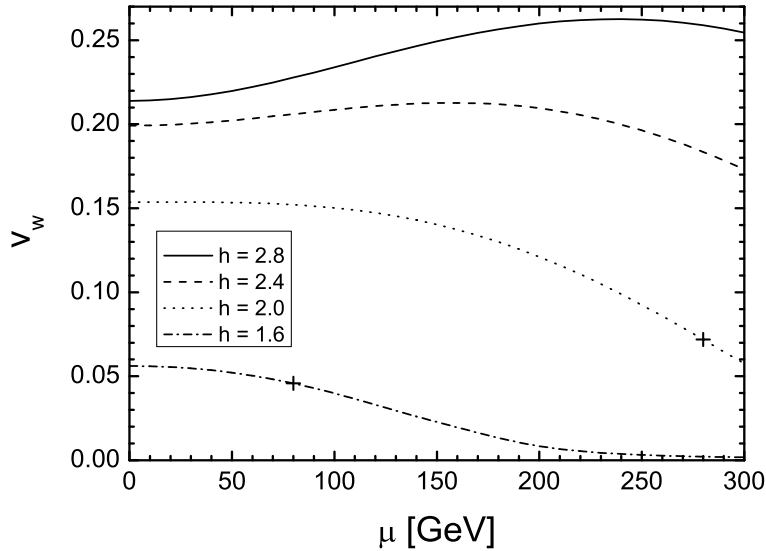}
\caption{The wall velocity as a function of $h$, for $g=12$ and $m_H=125GeV$.
For the points to the right of the crosses the phase transition is weak ($
\protect\phi_n/T_n<1$).}
\label{figfermu}
\end{figure}

In Fig. \ref{figfer} we have plotted the wall velocity for $\mu
_{f}=0$ and different values of $m_{H}$. The parts of the
curves on the right of the crosses correspond to $ \phi \left(
T_{n}\right) /T_{n}>1$. To obtain a strongly first-order phase
transition with extra fermions, large values of the Yukawa
coupling $h$ are needed. Therefore, the friction coefficient is
in general larger than in models in which the strength of the
transition is enhanced with extra bosons. As a consequence, the
wall velocity is smaller.  We find velocities $v_{w}\lesssim
0.25$, and as small as $v_w=0.05$ for strongly first-order
phase transitions. This makes this model interesting for
baryogenesis, since the generated baryon asymmetry peaks for
$v_{w}\ll 1$ \cite{lmt92,ckn92}.
Figure \ref{figfermu} shows the case $\mu \neq 0$, for
$m_{H}=125GeV$ and different values of $h$. In the lower
curves, the phase transition is weak on the right of the
crosses. We see that the wall velocity can either grow or
decrease as the invariant mass of the extra particles is
increased. Nevertheless, $v_{w}$ is still in the range
$0.05\lesssim v_{w}\lesssim 0.25$ for strongly first-order
transitions.

\subsection{Other extensions}

Given the large variety of extensions of the SM, it is not
feasible to consider all the models. For many models, the
strength of the phase transition depends only on the particle
content. Generally, adding bosons to the SM makes the phase
transition more strongly first-order, whereas fermions make it
weaker. To one-loop order, the magnitude of this effect depends
on the coupling of the extra particles to the Higgs boson, as
well as on their number of degrees of freedom. The friction
also depends on these parameters, increasing with the addition
of either bosons or fermions. The previous examples correspond
to such models, in which the extra-SM sector was dominated
either by bosons or by fermions. For an intermediate model in
which bosons and fermions have the same values of $g$, $h$, and
$\mu $, we have obtained intermediate values of the velocity,
namely, only deflagrations with $v_{w}\sim 0.15-0.35$ for phase
transitions with $\phi _{n}/T_{n}>1$.

On the other hand, the strength of the phase transition may be
changed without directly changing the particle content of the
model. This is what happens, e.g., in the MSSM when the
two-loop contribution (\ref{twoloop}) is taken into account.
Another possibility is the introduction of a non-renormalizable
dimension-six term of the form $\left( \Phi ^{\dagger }\Phi
-v^{2}/2\right) ^{3}/\Lambda ^{2}$ in the Higgs potential,
which allows to consider a negative quartic coupling
\cite{sixth}. Adding a real singlet field $S$ to the SM allows
for the possibility of cubic terms of the form $\left(
H^{\dagger }H\right) S$ or $S^{3}$ in the tree-level potential,
which cannot be constructed with Higgs doublets. The presence
of cubic terms already at zero temperature makes it easier to
get a strongly first-order electroweak phase transition
\cite{cv93}. This possibility exists also in the Next to
Minimal Supersymmetric Standard Model (NMSSM), which consists
of adding a gauge singlet to the MSSM. In this model the cubic
terms arise as supersymmetry-breaking soft terms. Since the
strength of the transition is dominated by the cubic terms in
the tree-level potential, it is not necessary to rely on loop
corrections or to consider a light stop \cite{nmssm}.

In these modifications of the SM, the phase transition can be
made strongly first-order without increasing the friction. In
order to explore the effect of such a tree-level modification
on the wall velocity, we shall consider the addition of cubic
terms. Considering the full potential makes the model
considerably more complicated than those we have studied so
far, since one has to deal with more than one scalar and
several free parameters. Instead, we shall consider a toy model
which consists of adding a term $-A\phi ^{3}$ to the tree-level
potential (\ref{v0}) for the SM, where $A$ is a free parameter
with mass dimensions. In this model the parameters of the
potential are related to the physical Higgs VEV and mass by
$2m^{2}=\lambda v^{2}-3Av$, $m_{H}^{2}=2\lambda v^{2}-3Av$. We
shall use the high-temperature potential
\begin{equation}
V\left( \phi ,T\right) =D_{SM}\left( T^{2}-T_{0}^{2}\right) \phi ^{2}-\left(
TE_{SM}+A\right) \phi ^{3}+\frac{\lambda }{4}\phi ^{4},  \label{potcubic}
\end{equation}
with the SM values given by $D_{SM}=\left(
2h_{W}^{2}+2h_{t}^{2}+h_{Z}^{2}\right) /8$, $E_{SM}=\left(
2h_{W}^{3}+h_{Z}^{3}\right) /6\pi $, and $T_{0}^{2}=m^{2}/D_{SM}$.

\begin{figure}[hbt]
\centering \epsfxsize=10cm \leavevmode \epsfbox{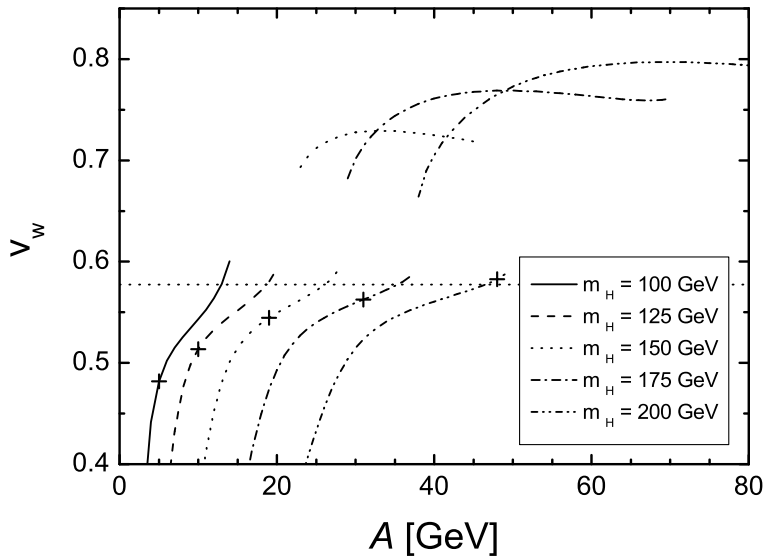}
\caption{The wall velocity for the SM with a cubic term $-A\protect\phi^3$.}
\label{figcubic}
\end{figure}

We show the results in Fig. \ref{figcubic}. We have considered
values of the parameters for which $h_{t}\phi /T\lesssim 1$, so
that the high-temperature expansion (\ref{potcubic}) is valid.
The strength of the phase transition increases very quickly
with $A$, and this is reflected in the behavior of the wall
velocity. Notice that, although in this case the particle
content does not change, the friction coefficient increases
with the strength of the transition due to its dependence on
the surface tension. Eventually, the wall velocity stops
growing. There is also a strong dependence on $m_{H}$. For
smaller values of $m_{H}$ ($m_{H}=100$ and $200GeV$) the
strengthening with $A$ is so fast that the deflagration
solution quickly becomes supersonic and disappears. In these
cases, the detonation solution does not appear before the phase
transition ceases to occur. Interestingly, detonations appear
for weaker phase transitions (higher values of $m_{H}$).

\section{Gravitational waves and baryogenesis}
\label{conseq}

The generation of gravitational waves and of the baryon
asymmetry of the Universe in the electroweak phase transition
seem to be mutually exclusive, since baryogenesis requires
small wall velocities while GW production requires large
velocities. In this section we discuss which of them is more
likely in the various extensions of the SM we have considered.

Both detonations and  deflagrations generate gravitational
waves
\cite{kkt94,kmk02,dgn02,cd06,kkgm08,hk08,amnr02,n04,gs07,cds08,m08}.
However, higher velocities give stronger signals, and Jouguet
detonations are often assumed for studying GW generation. As
shown in Ref. \cite{ms09}, the Jouguet velocity is not a good
approximation and detonations, when they exist, are weak
detonations. Furthermore, as we have seen, in the case of the
electroweak phase transition detonations exist only in some of
the extensions of the SM, and for extreme values of the
parameters, namely, small Higgs masses and large coupling $h$.
(Interestingly, a non-vanishing mass parameter $\mu$ may favor
the appearance of detonations, as can be seen in Fig.
\ref{figsing}). In any case, detonations are more likely in
models in which the strength of the transition is enhanced by
tree-level effects, as shown in Fig. \ref{figcubic}. On the
other hand, deflagrations with $v_w\gtrsim 0.1$ may also
produce a signal of GWs observable by LISA \cite{m08}. We
remark that the usual assumption that detonations are a
stronger source of GWs than deflagrations may be wrong. Weak
detonations have larger velocities than Jouguet detonations,
but cause a smaller disturbance in the fluid. On the other
hand, deflagrations with velocities close to $c_s$ may cause
important perturbations. For most of the models we have
considered, there is a wide range of parameters for which the
deflagration velocity is quite large. Extensions of the SM with
strongly coupled fermions constitute the only exception.

For the baryon asymmetry of the Universe to be produced at the
electroweak phase transition, the latter must be \emph{strong
enough}, i.e., the condition $\phi/T>1$ must be fulfilled. On
the other hand, the generated BAU is maximized for velocities
$v_w \lesssim 10^{-1}$, and is strongly suppressed for higher
velocities. This condition demands a \emph{weak enough} phase
transition, how weak depending on the friction. These two
conditions may restrict significantly the parameters of the
theory. Models with extra bosons which are strongly coupled to
the Higgs have been extensively considered, since they easily
give strongly first-order phase transitions. However, in our
examples we have seen that these models also tend to give quite
large wall velocities. For extensions with scalar singlets, we
obtained in all the cases $v_w\gtrsim 0.2$ for $\phi_n/T_n\geq
1$, and even larger ($v_w\gtrsim 0.5$) if tree-level cubic
terms are allowed. For the MSSM with a light stop we also
obtained rather large velocities, $v_w\gtrsim 0.35$. On the
other hand, extensions with fermions are not usually taken into
account because they tend to weaken the phase transition.
However, models with fermions can give strong enough phase
transitions, as shown in Ref. \cite{cmqw05} for the case of
fermions with large Yukawa coupling $h$ and heavier stabilizing
bosons. For this model, with $\mu_S$ given by Eq. (\ref{mus}),
we see that the wall velocity can be as small as $v_w=5\times
10^{-2} $ for $\phi_n/T_n=1$ (see Figs. \ref{figfer} and
\ref{figfermu}).

Due to the approximations used in the calculation of the
friction coefficient, the error in the wall velocity is an
$\mathcal{O}(1)$ factor. With a larger friction, extensions
with bosons may  give velocities $v_w\lesssim 0.1$ for
$\phi_n/T_n=1$. In any case, it is clear that values of
$\phi_n/T_n$ and $v_w$ which are appropriate for baryogenesis
are more easily obtained in models in which the dynamics is
dominated by fermions. If the friction is smaller than our
estimate, we will have higher velocities, which favors the
generation of GWs. For instance, in the MSSM, if we consider a
friction coefficient a factor of $3$ smaller we obtain
deflagrations with $v_w\approx c_s$ and detonations with $
v_w\approx 0.85$ for $m_H=120GeV$.

\section{Conclusions}
\label{conclu}

We have studied the velocity of bubble walls in the electroweak
phase transition. We have estimated the friction on the wall
due to particles with dispersion relation
$m^2=\mu^2+h^2\phi^2$, for arbitrary values of the parameters
$\mu$ and $h$. We have discussed analytically the cases of
small and large $\mu/T$ and $h\phi/T$. We have solved
numerically the equations for the wall velocity, taking into
account the friction and the hydrodynamics, and  we have
computed the electroweak wall velocity for several extensions
of the Standard Model. We have also discussed the implications
of our results for baryogenesis and gravitational wave
generation.

The friction coefficient, as well as the amount of
supercooling, have a strong, nontrivial dependence on the
coupling $h$ of the particles to the Higgs. Furthermore, the
friction depends on the bubble wall profile, and thus on the
strength of the phase transition. Therefore, the wall velocity
is not related to the strength of the transition in a simple
way, and the behavior depends on the model. We have found that
in general the velocity increases with the coupling $h$, but it
can also decrease (see, e.g., Fig. \ref{figsing}). As we
increase the mass parameter $\mu$, the friction tends to
increase for small $\mu$ and to decrease for large $\mu$.

As we have seen, detonations exist only for high values of $h$
and low values of $m_H$, except in the case in which the
strength of the phase transition is due to tree-level effects
rather than to loop contributions of extra particles. We stress
that there may be $\mathcal{O}(1)$ errors in our estimate of
the friction coefficient. For non-relativistic velocities this
will not introduce qualitative differences. However, if a model
has deflagrations with velocity close to the speed of sound, a
lower friction may cause the detonation solution to exist and
the deflagration to disappear. In particular, the interaction
rates $\Gamma $ need to be calculated accurately in each model.

We have also seen that it is difficult to obtain small
velocities in phase transitions with  $\phi_n/T_n\geq 1$. For
instance, the MSSM gives velocities in the range $0.35\lesssim
v_w\lesssim 0.45$, a hidden sector of scalar singlets gives
velocities ranging from $v_w\gtrsim 0.1$ up to supersonic
values, and a tree-level cubic term causes velocities
$v_w\gtrsim 0.5$. In these models the wall velocity is rather
large for baryogenesis. On the other hand, such high velocities
(either detonations or deflagrations) may produce an observable
signal of GWs. In contrast, the presence of fermions causes
smaller velocities, since they increase the friction without
increasing the phase transition strength. Thus, baryogenesis is
more likely in models with fermions. In particular, for the
model considered in Ref. \cite{cmqw05}, which contains fermions
with large Yukawa couplings $h$, we have found velocities in
the range $0.05\lesssim v_w\lesssim 0.25$ for $\phi_n/T_n\geq
1$.

\section*{Acknowledgements}

This work was supported in part by Universidad Nacional de Mar
del Plata, Argentina, grants  EXA 365/07 and 425/08. The work
by A.D.S. was supported by CONICET through project PIP 5072.
The work by A.M. was supported by CONICET through project PIP
112-200801-00943.

\end{document}